\documentclass[preprint,12pt]{elsarticle}
\usepackage{subfigure}
\usepackage[utf8]{inputenc}  
\usepackage[T1]{fontenc}  

\usepackage{amssymb}
\usepackage{amsmath}
\usepackage{lineno}

\journal{Nuclear Instruments and Methods in Physics Research Section A}

\begin{document}

\begin{frontmatter}



\title{Preliminary design and simulation for CEPC fast luminosity monitor detector based on 4H-SiC}


\author[1,2]{Yanpeng Li}
\author[2,3]{Meng Li}
\author[1]{Xingrui Wang}
\author[1]{Weimin Song \corref{cor1}}
\author[2]{Xiyuan Zhang \corref{cor1}}
\author[2]{Congcong Wang}
\author[4]{Suyu Xiao}
\author[2]{Haoyu Shi}
\author[2]{Dou Wang}
\author[3]{Philip Bambade}
\author[2]{Xin Shi}

\affiliation[1]{organization={College of Physics, Jilin University},
            addressline={2699 Qianjin Avenue, Chaoyang District}, 
            city={Changchun},
            postcode={130015}, 
            state = {Jilin},
            country={China}}

\affiliation[2]{organization={Institute of High Energy Physics, Chinese Academy of Sciences},
            addressline={19B Yuquan Road, Shijingshan District}, 
            city={Beijing},
            postcode={100049},
            country={China}}

\affiliation[3]{organization={{Laboratoire de Physique des 2 infinis Ir\'{e}ne Joliot-Curie– IJCLab}},
addressline={{B\^{a}t. 100, 15 rue Georges Cl\'{e}menceau}},
city={Orsay cedex},
postcode={91405},
country={France}}

\affiliation[4]{organization={Shandong Institute of Advanced Technology},
            addressline={NO.1501, Panlong Road}, 
            city={Jinan},
            postcode={250100}, 
            state = {Shandong},
            country={China}}

\cortext[cor1]{Corresponding author:
E-mail address: zhangxiyuan@ihep.ac.cn (Xiyuan\hspace{1ex}Zhang),\hspace{1ex}weiminsong@jlu.edu.cn (Weimin Song)}

\begin{abstract}
The Circular Electron-Positron Collider (CEPC), a next-generation high-luminosity collider, employs a crab waist scheme to achieve ultrahigh $5 \times 10^{34} \, \text{cm}^{-2}\text{s}^{-1}$ luminosity at Higgs mode. Owing to the extremely small beam size, the luminosity is highly sensitive to the stability of final focusing elements, where mechanical vibrations (e.g. ground motion) may induce beam offsets and luminosity degradation. To address this, a luminosity-driven dithering system is implemented for horizontal beam stabilization. In this work, we develop an optimized 4H-SiC fast luminosity detector scheme using an array of radiation detectors with picosecond time resolution positioned at critical locations. By using self-development software RAdiation SEmiconductoR (RASER), we optimize the active area of the detector to achieve 2\% relative precision at 1~kHz. Furthermore, the Total Sample Current (TSC) exhibits a near-linear correlation with luminosity attenuation, enabling real-time luminosity monitoring.
\end{abstract}




\end{frontmatter}



\section{Introduction} \label{sec:introduction}

CEPC is a 100~km double-ring collider featuring two interaction points (IPs). It is designed to operate under four center-of-mass energy configurations: Z mode (91~GeV), W mode (160~GeV), Higgs mode (240~GeV), and top quark pair production ($t\bar{t}$) mode (360~GeV). Its primary focus will be the Higgs mode, aiming to precisely measure Higgs boson properties and rigorously test the Standard Model. The target nominal luminosity for the Higgs mode is $5 \times 10^{34} \, \text{cm}^{-2}\text{s}^{-1}$\cite{Gao2024}. To achieve this desired luminosity target, CEPC intends to implement the crab waist scheme employed by SuperKEKB\cite{PhysRevAccelBeams.19.121005, AKAI2018188}. The CEPC design specifies horizontal and vertical beam sizes of 14~$\mu$m and 36~nm, respectively, at IP. However, maintaining precise alignment between the positron and electron beams within such a small tolerance is challenging. Thus, real-time beam orbit monitoring and calibration are required. In the vertical direction, the Beam Position Monitor (BPM) serves a crucial function. However, due to the limitations of the BPMs resolution in the horizontal direction, a luminosity-driven dithering orbit feedback system is employed, similar to the approach used by SuperKEKB[\citenum{pang:tel-03092297}, \citenum{PANG2019225} and references therein].

The luminosity-driven dithering orbit feedback system monitors the luminosity - and is therefore sensitive to the offset of the orbit - by detecting the number of the radiative Bhabha events at the vanishing photon angle. Electrons scattered in the radiative Bhabha process at the IP, referred to as primary electrons, undergo varying degrees of trajectory deflection due to the magnetic fields as they propagate downstream from the IP, depending on their energy. As a result, primary electrons with different energies impact the beam pipe at different positions. This interaction subsequently triggers electromagnetic showers, producing secondary particles. A detector placed outside the beam pipe captures these secondary particles and measures the radiative Bhabha electron count based on the detector's signal response.

Silicon carbide (SiC), especially 4H-SiC, has a stronger radiation tolerance compared to silicon. In previous studies \cite{RUDDY2007163,1710303,9217477}, 4H-SiC has shown excellent performance in various harsh radiation environments. Under photon irradiation with a dose of 5.4~MGy, electron irradiation at 2~MeV with a fluence of $1 \times 10^{16}\,\mathrm{electrons/cm^2} $, neutron irradiation at 1~MeV with a fluence of $8 \times 10^{15}\,\mathrm{neutrons/cm^2} $, and 24~GeV proton irradiation with a fluence of $2.5 \times 10^{15}\,\mathrm{protons/cm^2} $, the 4H-SiC detectors maintained excellent signal response to alpha particles. In addition, the lower leakage current and higher thermal conductivity of 4H-SiC enable it to operate at room temperature, and even up to 500~°C\cite{GAL2023157708}, without the need for a cooling system, saving space inside the tunnel. Furthermore, the 4H-SiC detector demonstrates exceptional time resolution. Our group reported that the 5~mm $\times$ 5~mm PIN-type detector constructed from 4H-SiC, achieves an outstanding time resolution of $94 \pm 1$~ps\cite{10.3389/fphy.2022.718071}. In comparison to diamond detectors, SiC detectors are more cost-effective and better suited for large-scale applications. Due to these properties, SiC has been used for $\mu$-beam monitor for the COMET experiment at J-PARC\cite{Kishishita_2025}.

In this study, we utilized RASER\cite{RASER} to simulate the interaction between primary electrons and beam pipes, enabling us to determine the distribution of secondary particles. By employing the Shockley-Ramo theorem\cite{10.1063/1.1710367, 1686997}, we simulated the current signals produced by these secondary particles within the detector. The detector geometry was subsequently optimized to achieve a relative precision of 2\% under 1~kHz operational conditions, accounting for ground motion (GM) effects\cite{Pang:IPAC2018-WEPAL037}. The resulting detector geometry ensures compatibility with the nominal luminosity specifications of the Higgs operation mode. Additionally, the response of the detector to the decrease in luminosity was also simulated. The simulation results show that the derived Total Sample Current (TSC) values within 1~ms exhibit a good linear relationship with luminosity, demonstrating the suitability  for real-time luminosity monitoring.

\section{Primary electron tracking}\label{sec: Primary electrons tracking}

To detect a sufficient number of events for achieving the target precision, the monitoring system for the low-energy radiative Bhabha-scattered electrons should preferentially select locations with both higher beam loss rates and concentrated distributions, permitting the implementation of a more compact detector system. Through simulations combining BBBREM\cite{KLEISS1994372} for the radiative Bhabha events at the vanishing photon angle and SAD\cite{website:sad} for electron trajectory tracking up to 100~meters downstream of the IP,  three optimal detector locations were identified for meeting the CEPC spatial constraints\cite{Li2025}. Three candidate monitoring positions are located at 10~meters, 84~meters, and 90.5~meters downstream of IP, respectively. Position 2, situated in the gap between segments of the first bending magnet, presents substantial spatial constraints and mechanical interference issues, making it unsuitable for deployment. Both Position 1 at 10~meters and Position 3 at 90.5~meters are suitable as monitoring locations. Notably, at Position 1, lost electrons exhibit symmetric distribution on both sides of the beam pipe. Any horizontal offset between the two colliding beams at the IP will cause the beams to mutually focus each other through their attractive electromagnetic interaction, thereby deflecting each beam with respect to its original path. Electrons scattered through the radiative Bhabha process are also subjected to such deflections further disrupting this symmetry. It has been checked that by summing events from identical detectors located on both sides of Position 1, the dependence in event counting rate on horizontal beam offsets at the collision point can be largely mitigated. The basic information regarding the incident primary electrons at these two positions, along with the corresponding detector requirements, is presented in the Table~\ref{Tab:positions} below.

\begin{table}[h] 
    \centering 
    \resizebox{0.8\textwidth}{!}
    {
        \begin{tabular}{c c c c} 
            \hline 
            & Position 1  & Position 3\\ 
            \hline
            Distance from IP & 10~m & 90.5~m\\
            Average Number detected/collision & 3.4(two sides) & 3.2(one side) \\
            Average Number detected/ms & 2830 & 2670 \\
            Expected Relative Precision & 2\% @1~kHz & 2\% @1~kHz\\
            Average Energy of scattered electron & 24~GeV & 75.3~GeV\\
            Average Hitting Angle & $1.7 \times 10^{-4}$~rad & $7 \times 10^{-4}$~rad\\
            Primary Electron Losses Area & $5 \times 20 \times 2\, ~\text{cm}^{2}\ $ & $3 \times 15 \times 1\, ~\text{cm}^{2}\ $\\
            Detector Number & 2 & 1\\
            Detector Measurement Parameters & \multicolumn{2}{c}{Number of signals within 1 ms}\\
            Detector Time Resolution & \multicolumn{2}{c}{600~ns}\\
            \hline
        \end{tabular}
    }
    \caption{Basic information of the incident primary electrons and requirements for the detectors at two positions}
    \label{Tab:positions}
\end{table}

\begin{figure}[h]
    \centering
    \includegraphics[scale=0.25]{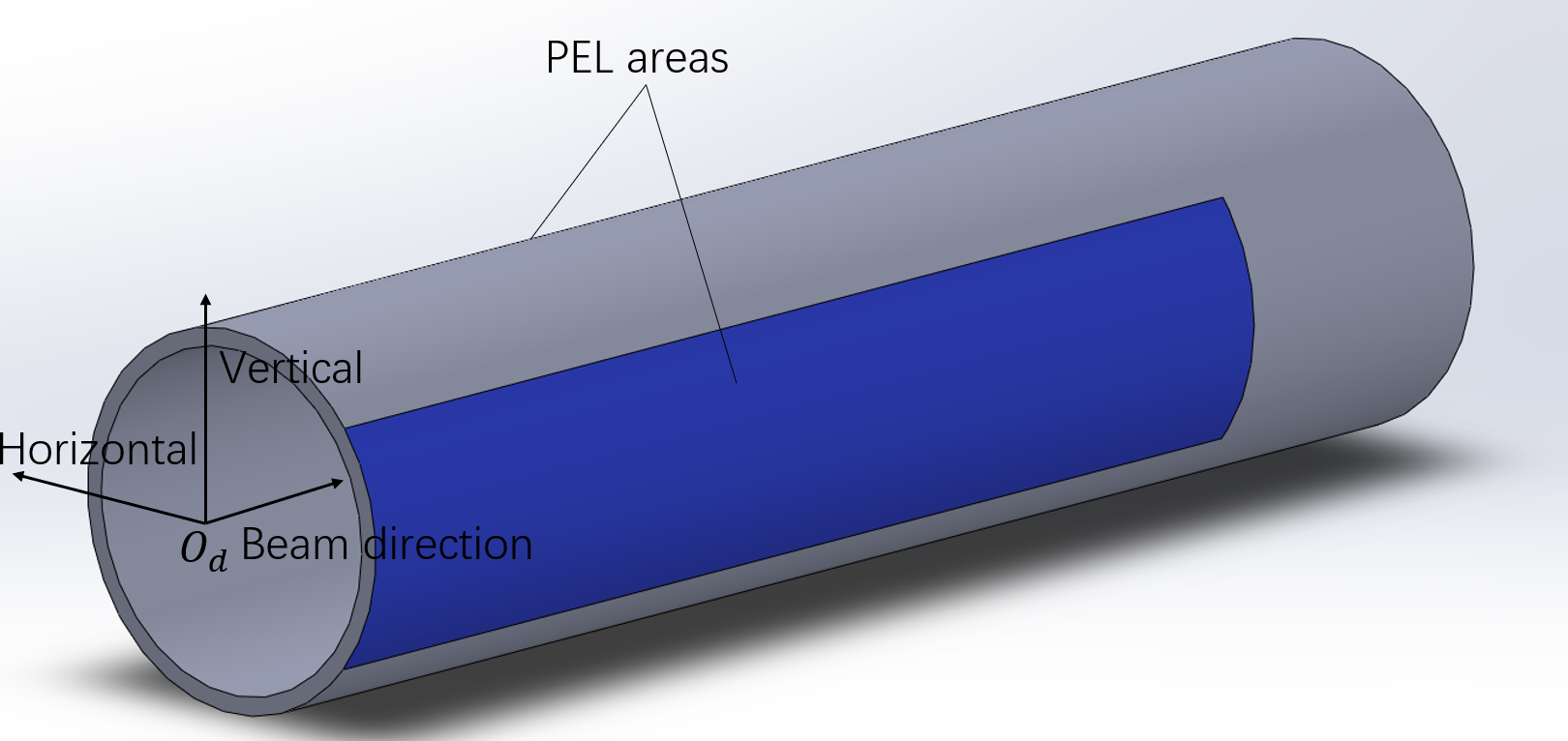}
    \caption{The illustration of the PEL area and the $O_{d}$ coordinate system at Position 1}
    \label{fig:PEL_area}
\end{figure}

\begin{figure}[h]
    \centering
    \subfigure[]{ \label{fig:primary_hit_XoY}
    \includegraphics[scale=0.3]{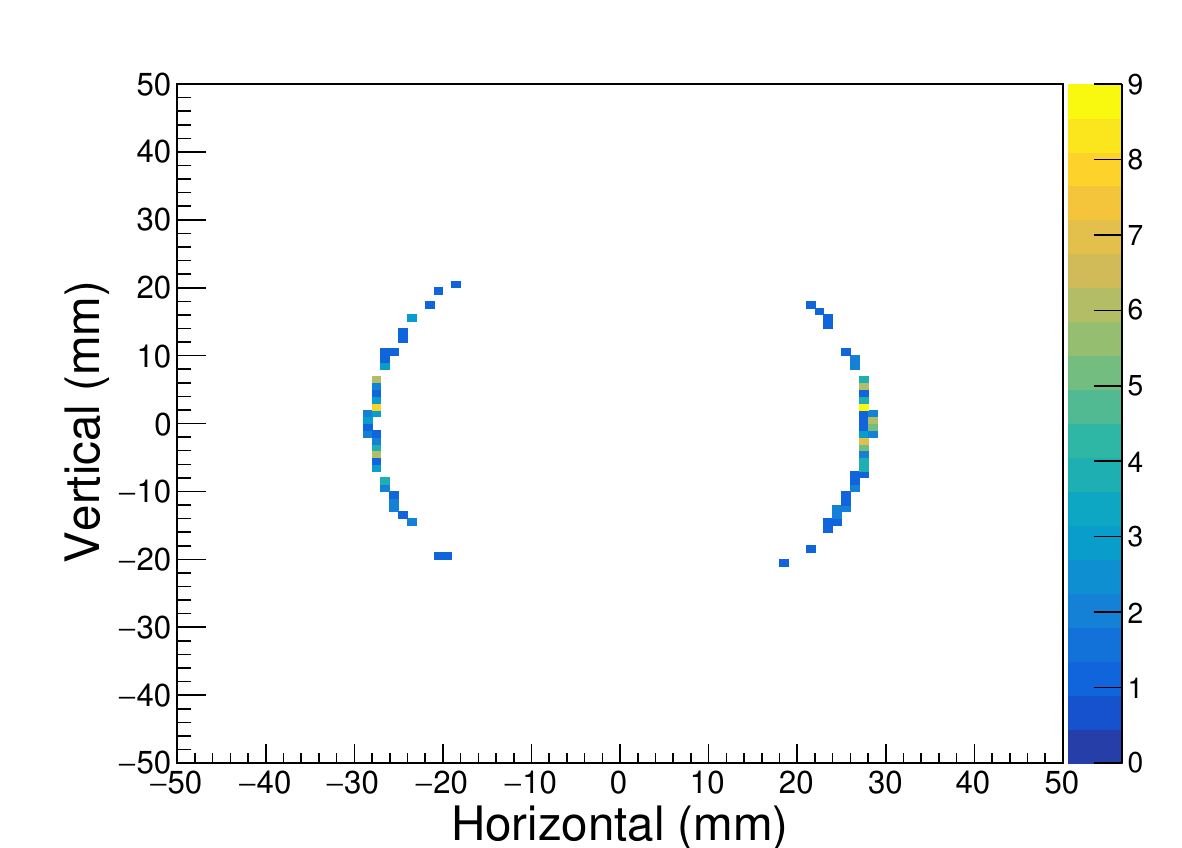}}  
    \subfigure[]{ \label{fig:primary_hit_Z}
    \includegraphics[scale=0.3 ]{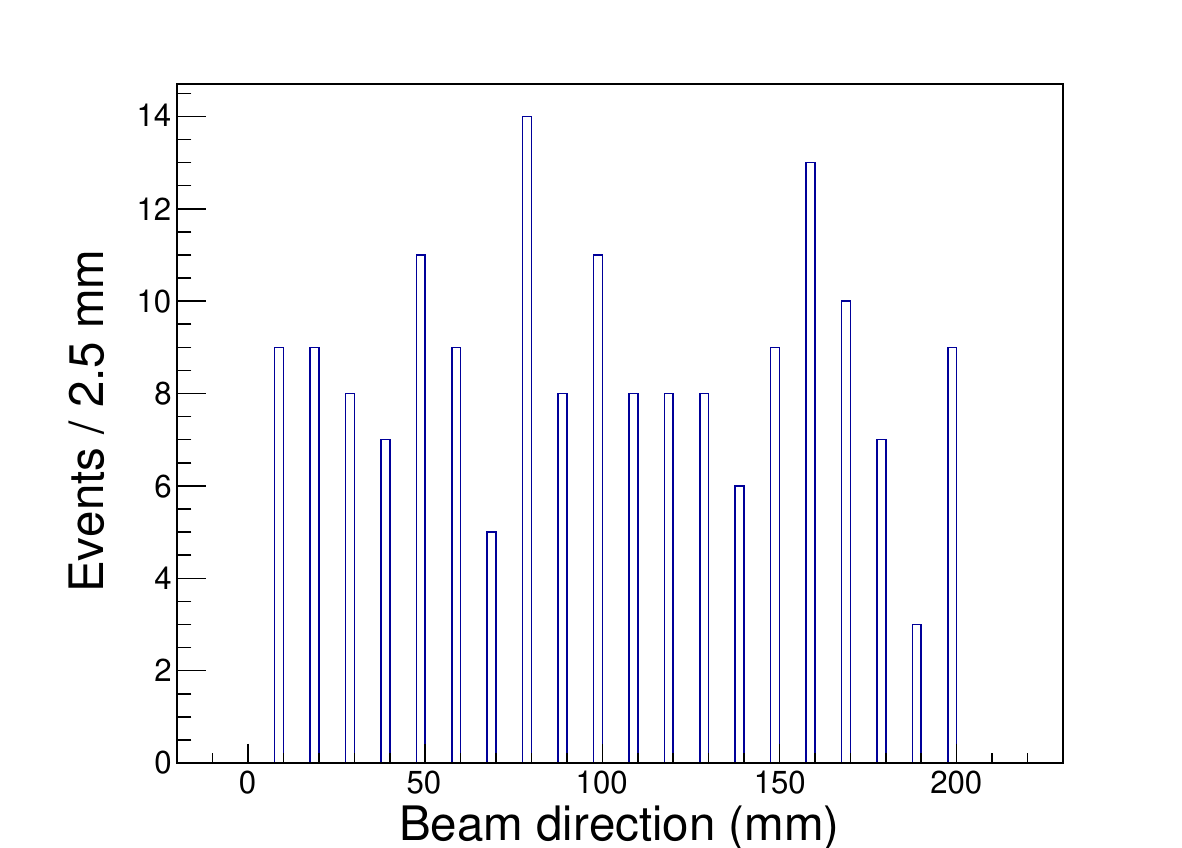}}
    \caption{Distribution of hit position at Position 1 within the $O_{d}$ coordinate: (a) The distribution of primary electrons at Horizontal-Vertical plane and (b) The distribution of primary electrons along Beam direction}
    \label{fig:position}
\end{figure}

In the following, we focus on a detector design plan for Position 1 along with the corresponding simulation results. At Position 1, which exhibits the highest density of primary electron impacts, primary electrons with an average energy of 24~GeV collide with the beam pipe at an average angle of $1.7\times 10^{-4}$~radians. In this first detailed study, two Primary Electron Losses(PEL) areas are determined on either side of the beam pipe in the horizontal direction. Each PEL area consists of curved surfaces with dimensions of 5~cm $\times$ 20~cm. The secondary particles produced by the primary electrons hitting these two areas must be accepted by the detector and generate the corresponding response. A coordinate system, denoted as $O_{d}$ at Position 1, is established as shown in Figure~\ref{fig:PEL_area}. Within the $O_{d}$ coordinate, the spatial distribution of PEL areas at Position 1 is shown in Figure~\ref{fig:position}, where Figure~\ref{fig:primary_hit_XoY} shows the distribution of primary electrons at the Horizontal-Vertical plane and Figure~\ref{fig:primary_hit_Z} shows the distribution of primary electrons along the beam direction. Figure~\ref{fig:position} shows the symmetry in the vertical distribution of primary electrons and their uniform distribution along the beam direction.

For high-luminosity electron-positron colliders, such as CEPC, the single beam loss background can impact the signal response of fast luminosity detectors. This background is primarily composed of beam-thermal photon scattering, beam-gas scattering including single Coulomb scattering and Bremsstrahlung scattering, and Touschek scattering\cite{Xu2022, Shi:2021+0}. Detailed simulation results indicate that these background processes at Position 1, situated 10~meters downstream of IP, are negligible in comparison to the signal processes. Consequently, the influence of the single beam loss background on the signal response of fast luminosity detectors was not accounted for in this study. Concerning synchrotron radiation, the critical energy of synchrotron radiation photons in Higgs mode is 30~keV. Simulation results indicate that photons at this energy interact with the beam pipe through Compton scattering and the photoelectric effect, remaining within the beam pipe. Additionally, the secondary electrons generated will remain in the beam pipe due to ionization energy loss and multiple scattering, thereby not affecting the detector. Moreover, the interaction between synchrotron radiation and the beam pipe can be effectively reduced by coating the inner wall of the beam pipe with gold, which further lowers the impact of synchrotron radiation on the detector. Therefore, the influence of synchrotron radiation is not considered in this work.

\section{Geometric design and simulation}

\subsection{Secondary particles tracking}

\begin{figure}[h]
    \centering
    \includegraphics[scale=0.25]{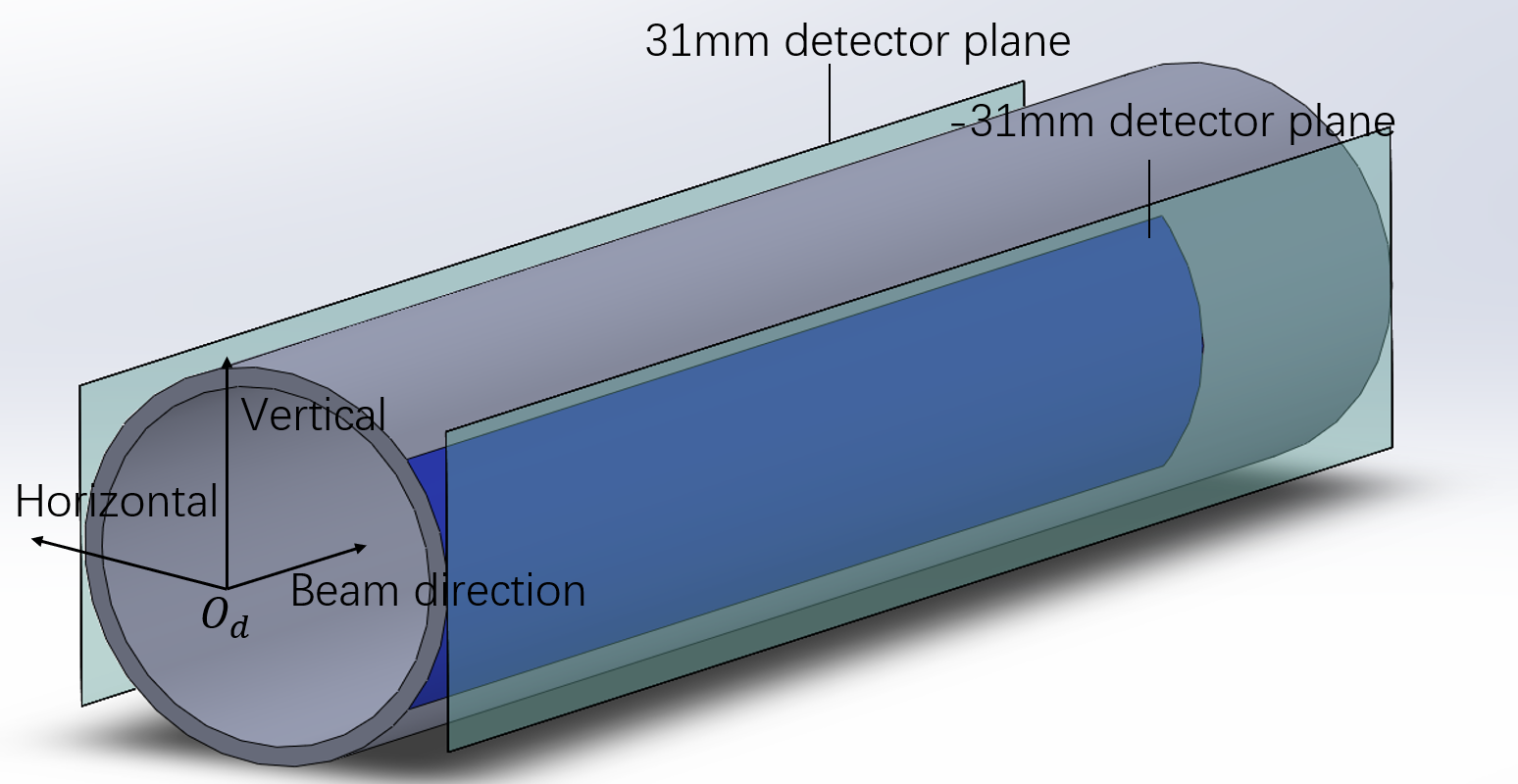}
    \centering
    \caption{The spatial relationship between the beam pipe and the detector plane}
    \label{fig:detplane}
\end{figure}

The detector performs monitoring by capturing secondary particles generated from the interaction between primary electrons and the beam pipe. To enhance detector response performance, it is essential to optimize both its spatial positioning and geometric dimensions for maximizing secondary particle collection efficiency. The beam pipe features an inner
diameter of 56~mm and an outer diameter of 62~mm (see \cite{Gao2024}). The detector plane - defined as the mounting surface for the detector - is positioned tangentially to the pipe's outer surface. Figure~\ref{fig:detplane} illustrates the spatial relationship between the beam pipe and the detector plane. Notably, the dimensions provided within the figure do not correspond to the actual sizes. In the simulation the two detector planes are set to a sufficiently large size (6~m $\times$ 6~m) to fully receive the secondary particles. Table~\ref{Tab:sp_species} presents the ratio of the number of various types of secondary particles, denoted as $N_{sp}$ to the total count of secondary particles, represented as $N_{total}$, detected the by two separate detector planes. The types and proportions of secondary particles received by the two detector planes are roughly the same, and photons and electrons account for the vast majority of the proportion. This category of "others" encompasses positrons, protons, and neutrons, among other particles. In the evaluation of detector radiation hardness, priority must be given to assessing the impact of photons on detector performance degradation mechanisms. SiC exhibits a high Si-C bond energy of up to 4.6~eV, twice that of the Si-Si bond ($\sim$2.3~eV) in silicon, significantly reducing the probability of atomic displacement induced by gamma irradiation. Additionally, the displacement threshold energy of SiC ($\sim$20–35~eV) is nearly double that of silicon (12–15~eV), making it more difficult for gamma rays to induce lattice defects. As a result, SiC demonstrates outstanding gamma radiation tolerance(see \cite{RUDDY2007163}) , making SiC detectors an ideal choice as fast luminosity detectors for the CEPC.

\begin{table}[h] 
    \centering 
    \resizebox{0.8\textwidth}{!}
    {
        \begin{tabular}{c c c} 
            \hline 
            & \multicolumn{2}{c}{$N_{sp}/N_{total}$(\%)}\\ 
            \hline
            Secondary particle species & -31~mm detector plane & 31~mm detector plane \\
            $\gamma$ & 87.86 & 87.67 \\
            $e^{-}$ & 7.19 & 7.32 \\
            others & 4.95 & 5.01\\
            total & 100 & 100\\
            \hline
        \end{tabular}
    }
    \caption{The ratio of the number of various secondary particle types (denoted as $N_{sp}$) to the total secondary particle count (denoted as $N_{total}$) detected by two independent detector planes}
    \label{Tab:sp_species}
\end{table}

\begin{figure}[h]
    \centering
    \subfigure[]{ \label{fig:spd_y}
    \includegraphics[scale=0.3]{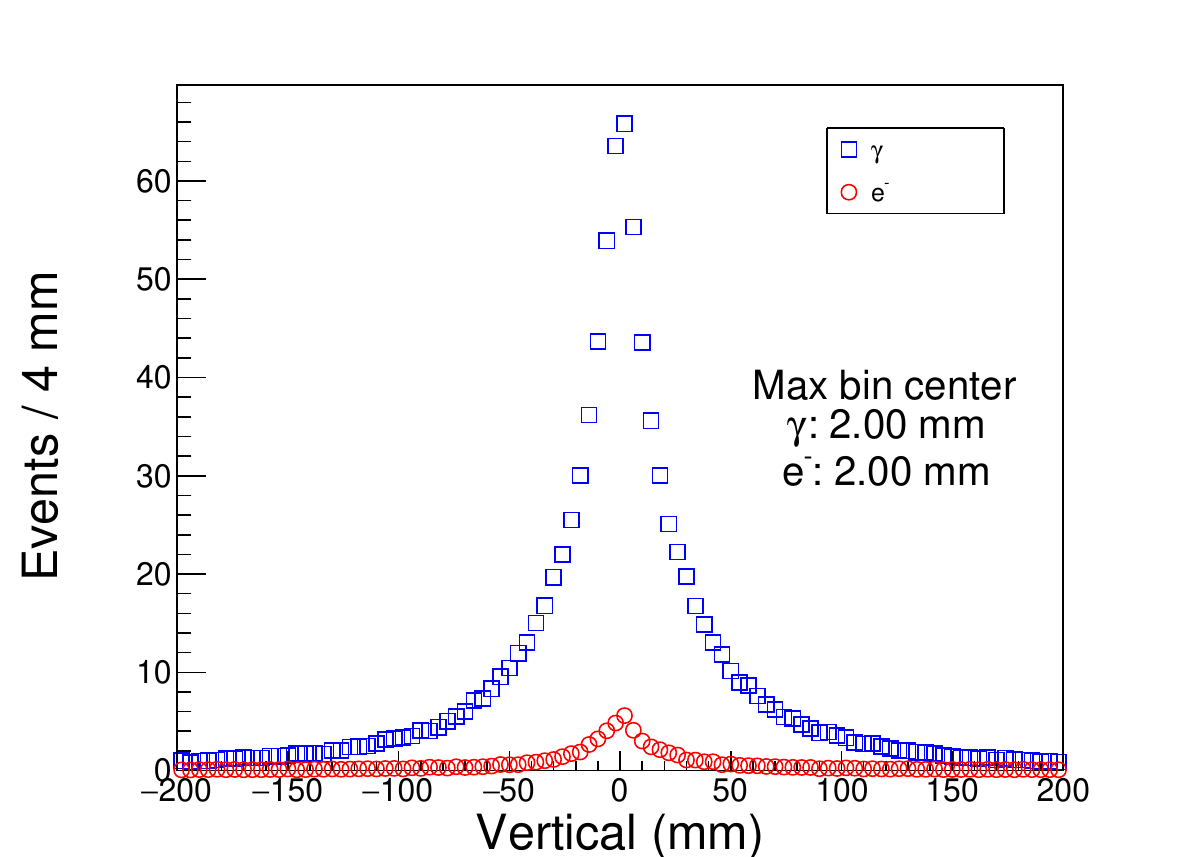}}
    \subfigure[]{ \label{fig:spd_z}
    \includegraphics[scale=0.3]{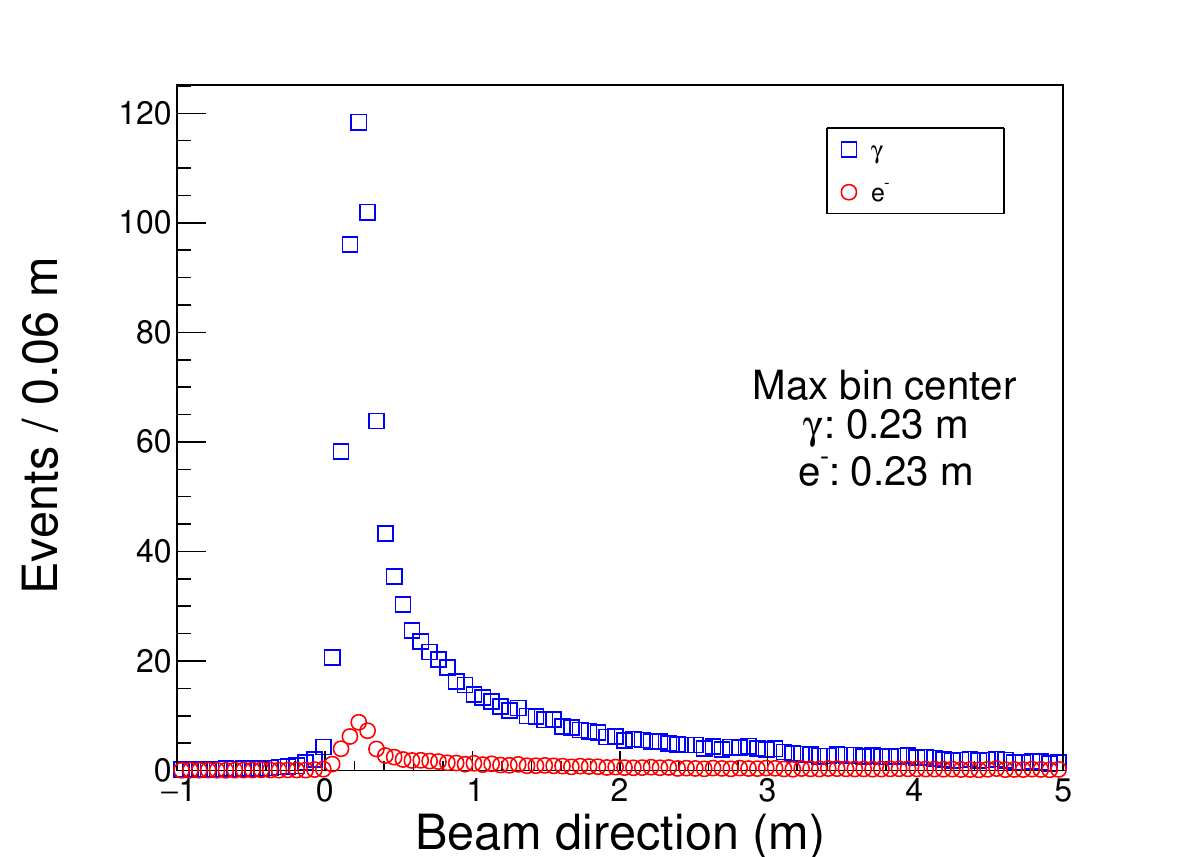}}
    \caption{At -31 mm detector plane in $O_{d}$ system, the vertical distribution (a) and  beam direction distribution (b) of secondary photons/electrons generated by per primary electron; in the vertical direction, the peak is located at 2~mm (near the origin), while in the beam direction, the peak appears at 23~cm.}
    \label{fig:spd_nx}
\end{figure}

Figure~\ref{fig:spd_y} shows the vertical distributions within the $O_{d}$ coordinate of secondary photons and electrons generated per primary electron incident on PEL areas striking the -31~mm detector plane. As illustrated, the secondary photons and electrons are concentrated near the origin and exhibit a symmetric distribution. Moreover, as shown in Figure~\ref{fig:spd_z}, in the beam direction, upon incidence of a primary electron on a beam pipe, an electromagnetic cascade is initiated, leading to a decrease in the electron energy. Subsequently, when the electron energy falls below a critical threshold, it dissipates its energy through ionization and excitation processes, rather than through the production of additional shower particles\cite{ParticleDataGroup:2024cfk}. A similar pattern is evident at the 31~mm detector plane; therefore, it is not displayed. Despite the secondary particles exhibiting extensive dispersion on both detector planes, which results in a weak correlation between the detector's signal response and its position, the central positions of the two detectors are maintained optimally at $P_{1} = (-31~mm, 0, 230~mm)$ and $P_{2} = (31~mm, 0, 230~mm)$ within the $O_{d}$ coordinate.

\subsection{Optimization of detector geometry}

In order to maximize the collection of secondary particles while taking into account the space constraints within the tunnel and the complexity of the readout electronics system, it is necessary to determine the optimal size of the entire detector. The 4H-SiC detector is composed of multiple pixel units. Each pixel has an area of 5~mm $\times$ 5~mm  with ohmic contacts on both top and bottom surfaces, featuring an active region with a doping concentration of $5.2 \times 10^{13} \,~\text{cm}^{-3}$ and a thickness of 100~$\mu$m (For more details, refer to ref\cite{10.3389/fphy.2022.718071}). As detailed in Section 3.1 the detection system was located at positions $P_{1}$ and $P_{2}$ corresponding to peak secondary particle fluxes ($\gamma$ and $e^{-}$). Particle distribution simulations along the beam direction demonstrate that a 1-cm-long detector can capture about 1.8\% of the secondary gammas and 1.6\% of the secondary electrons, establishing this as the standardized longitudinal dimension. To further optimize the detector's active area, a parametric scan was performed by systematically varying the vertical length (1-5~cm range) of the detectors installed on both sides of the beam pipe. In addition, due to the noise introduced by the detector and electronic systems, signals with smaller amplitudes may be overwhelmed by the noise. Therefore, when considering the relationship between the detector size and signal response, it is necessary to set current thresholds to describe the impact of the noise. Table~\ref{Tab:response of different size} presents the number of detected signal events within 1~ms across various detector vertical lengths at predetermined current thresholds, including 0 (used as a contrast), 0.05~$\mu$A, 0.5~$\mu$A, 1~$\mu$A, 3~$\mu$A and 5~$\mu$A. Following a bunch crossing event, if any detector on either side produces a current signal exceeding the predefined current threshold, it is deemed to have initiated a response.  From Table~\ref{Tab:response of different size}, it is evident that an increase in the detector's vertical dimension correlates with a corresponding increase in the number of detected signal events. Furthermore, within the same dimensions, an elevation in the set current threshold is associated with a decrease in the number of detected signal events.

\begin{table}[h] 
    \centering 
    \resizebox{0.8\textwidth}{!}
    {
        \begin{tabular}{c c c c c c} 
            \hline 
            Current threshold & 1 cm  & 2 cm & 3 cm & 4 cm & 5 cm\\ 
            \hline
            0           & 674 & 731 & 746 & 750 & 753\\
            0.05 $\mu$A & 673 & 731 & 745 & 750 & 753 \\
            0.50 $\mu$A & 672 & 731 & 743 & 750 & 753 \\
            1.00 $\mu$A & 653 & 717 & 738 & 743 & 750\\
            3.00 $\mu$A & 549 & 654 & 675 & 698 & 713\\
            5.00 $\mu$A & 484 & 600 & 619 & 638 & 673\\
            \hline
        \end{tabular}
    }
    \caption{The number of detected signal events with different vertical lengths at various current thresholds}
    \label{Tab:response of different size}
\end{table}

When the two beams are normally aligned, there are an average of 3.4 primary electrons hitting PEL areas per bunch crossing(see Table~\ref{Tab:positions}), so the number of radiative Bhabha electrons detected within 1~ms $N_{d}$ and the corresponding precision $\nu$ can be expressed as:

\begin{equation}\label{eq:(2)}
\begin{aligned}
& N_{d} = N_{0}\times\ \frac{N_{b}}{1-P(X=0)} \\
&P(X = k) = \frac{N_{0}^k}{k!} \times \exp(-N_{0}), \quad k = 0, 1, 2, \dots\\
& \nu =  \frac{1}{\sqrt{N}}
\end{aligned}
\end{equation}

\begin{figure}[h]
    \centering
    \subfigure[]{ \label{fig:np}
    \includegraphics[scale=0.3]{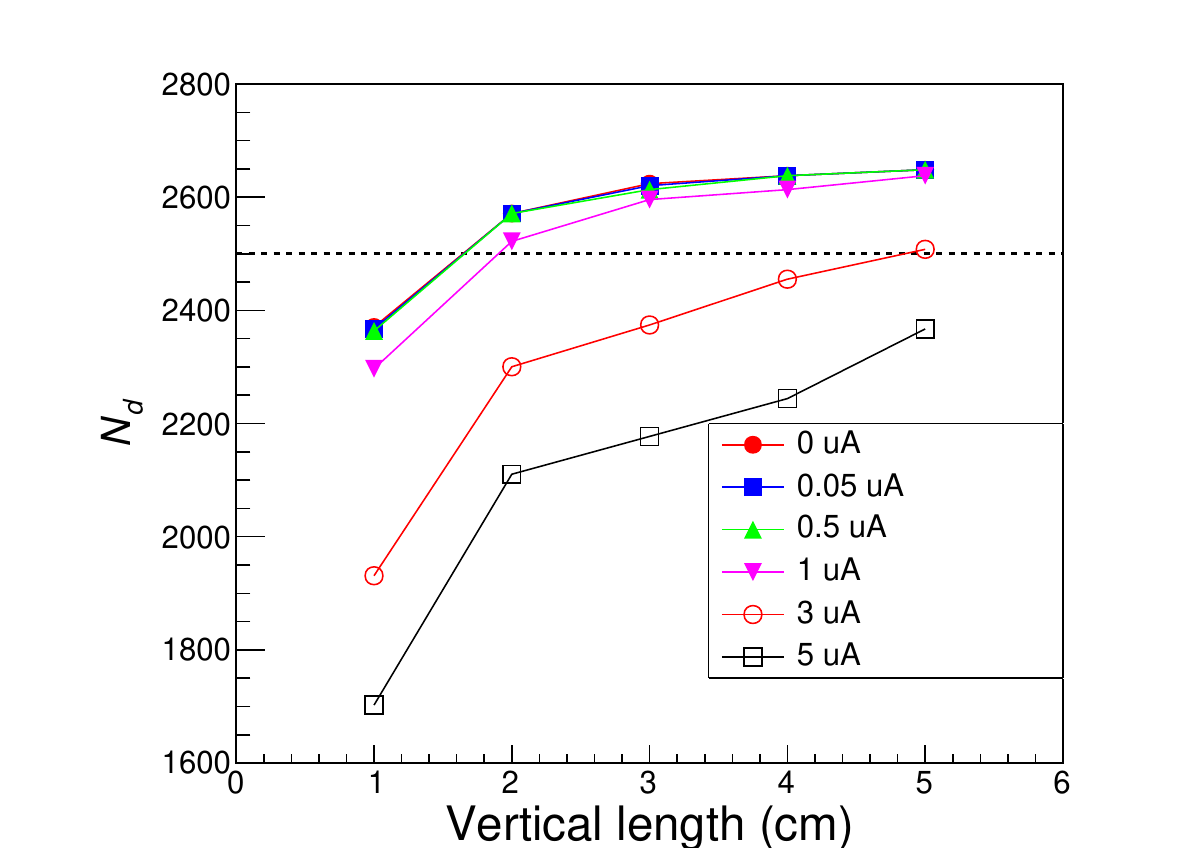}}  
    \subfigure[]{ \label{fig:precision}
    \includegraphics[scale=0.3]{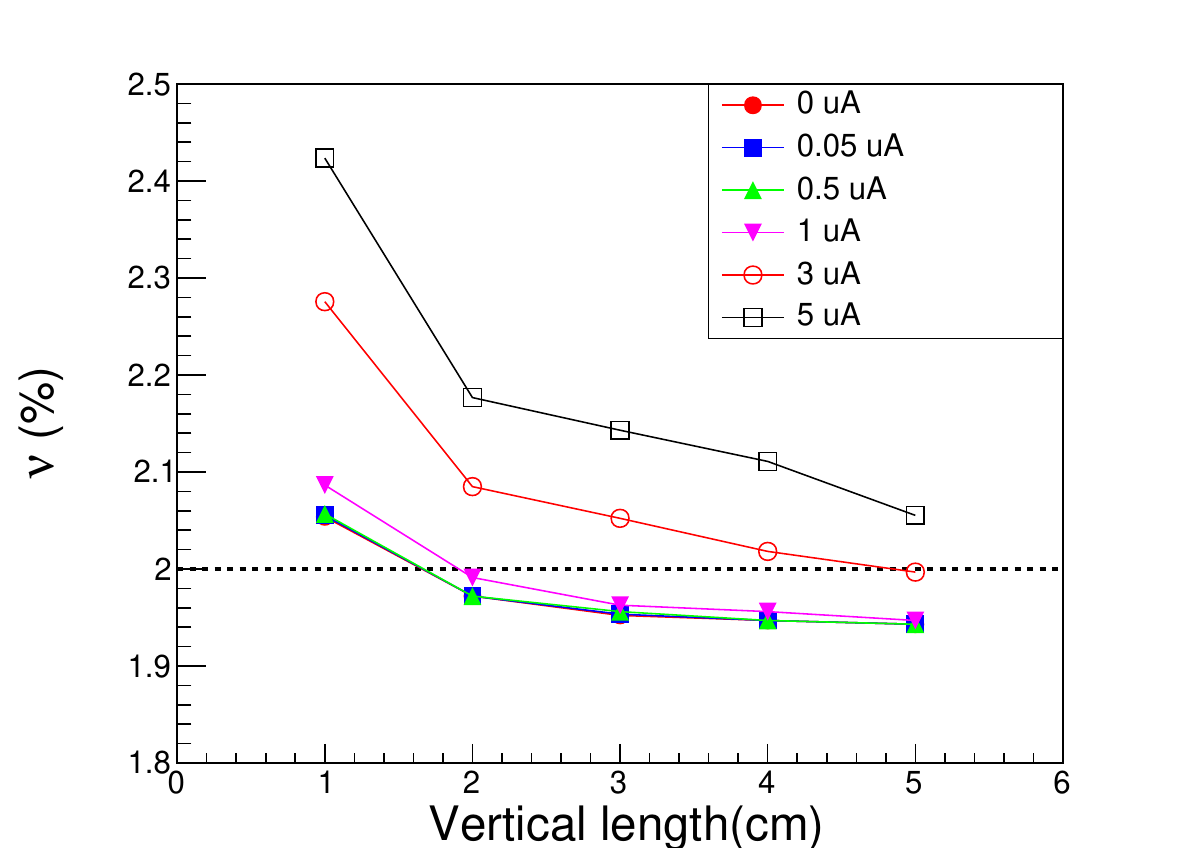}}
    \caption{(a) The number of radiative Bhabha electrons detected by detectors within 1 ms $N_{d}$ with varying vertical lengths at distinct current thresholds, along with (b) the associated precision $\nu$}
    \label{fig:np_precision}
\end{figure}

\begin{figure}[h]
    \centering
    \includegraphics[scale=0.21]{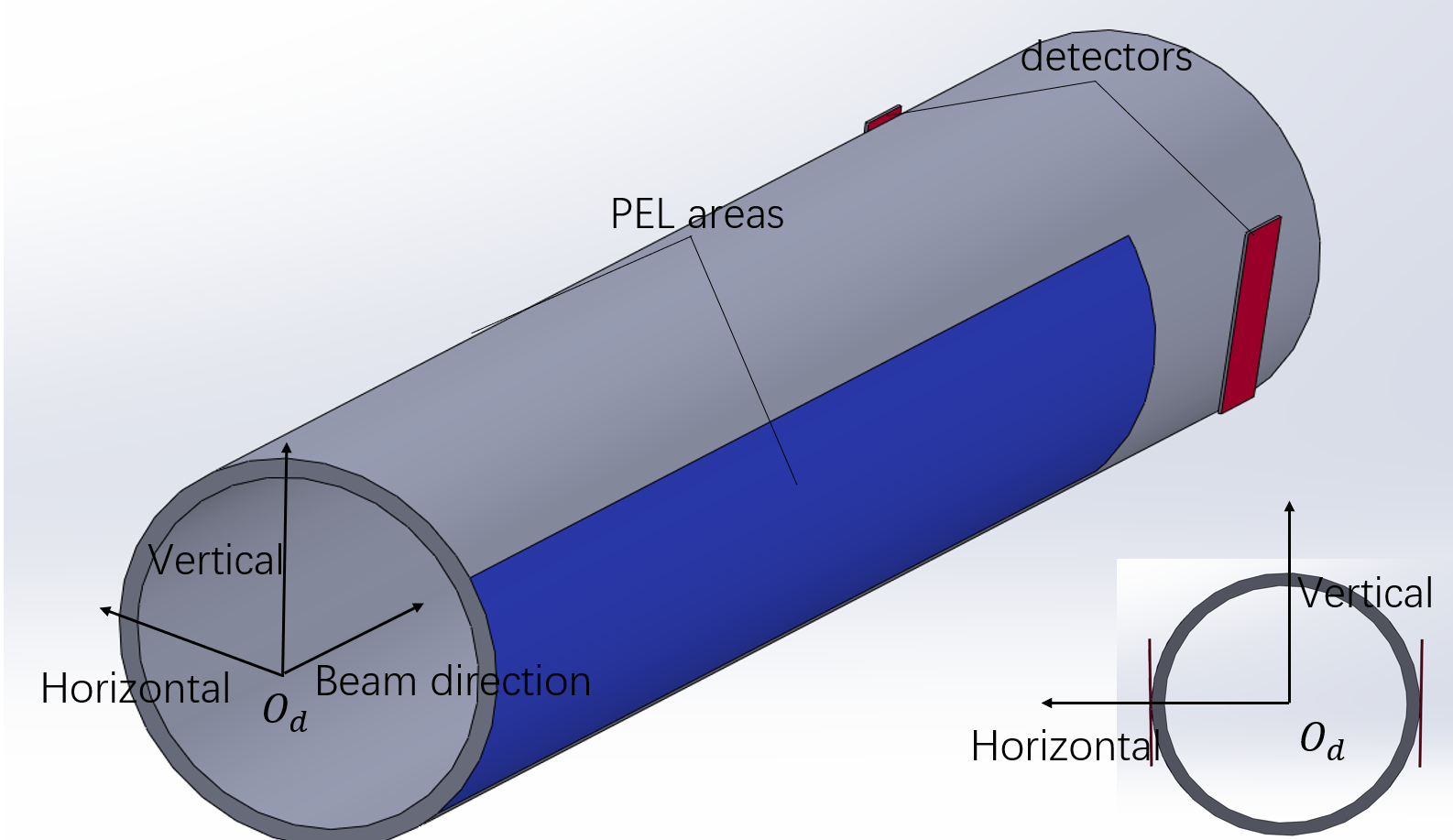}
    \caption{The schematic of the detector and beam pipe}
    \label{fig:schematic_detector_pipe}
\end{figure}

Among them, $N_{0}$ = 3.4 represents the number of particles that strike the PEL areas after each bunch crossing when the two beams are aligned in their normal configuration, while $N_{b}$ denotes the number of detected signal events within a 1~ms interval. $\nu$ is the relative precision of the primary electrons detected within 1~ms. The Poisson distribution was introduced because the number of primary electrons hitting the PEL areas produced by each bunch crossing follows a Poisson distribution with a mean of $N_{0}$. Figure~\ref{fig:np_precision} illustrates the number of radiative Bhabha electrons detected by detectors with varying vertical lengths at distinct current thresholds, along with the associated relative precision. The CEPC fast luminosity detector is required to achieve a precision of 2\% at 1~kHz, necessitating the detection of 2500 primary electrons within 1~ms. As illustrated in Figure~\ref{fig:np_precision}, smaller vertical dimensions (1~cm) of the detector and elevated current thresholds (3 $\mu$A and 5 $\mu$A) both contribute to a significant reduction in the number of detected primary electrons, consequently failing to satisfy the requisite precision criteria. Although a minimum vertical length of 2~cm is necessary to achieve this level of precision, a vertical length of 3 cm is implemented to guarantee compliance with the precision requirements. The detector is constructed from 12 pixel elements and corresponding 12 readout channels. The equivalent input current of noise in each channel is less than 1~$\mu$A. At the specified size and noise thresholds, the detection rate of primary electrons is 2594.3 per millisecond, with an associated precision of 2\%. Figure~\ref{fig:schematic_detector_pipe} presents the schematic of the detector and beam pipe.

\subsection{Detector performance simulation}

\begin{figure}[h]
    \centering
    \includegraphics[scale=0.30]{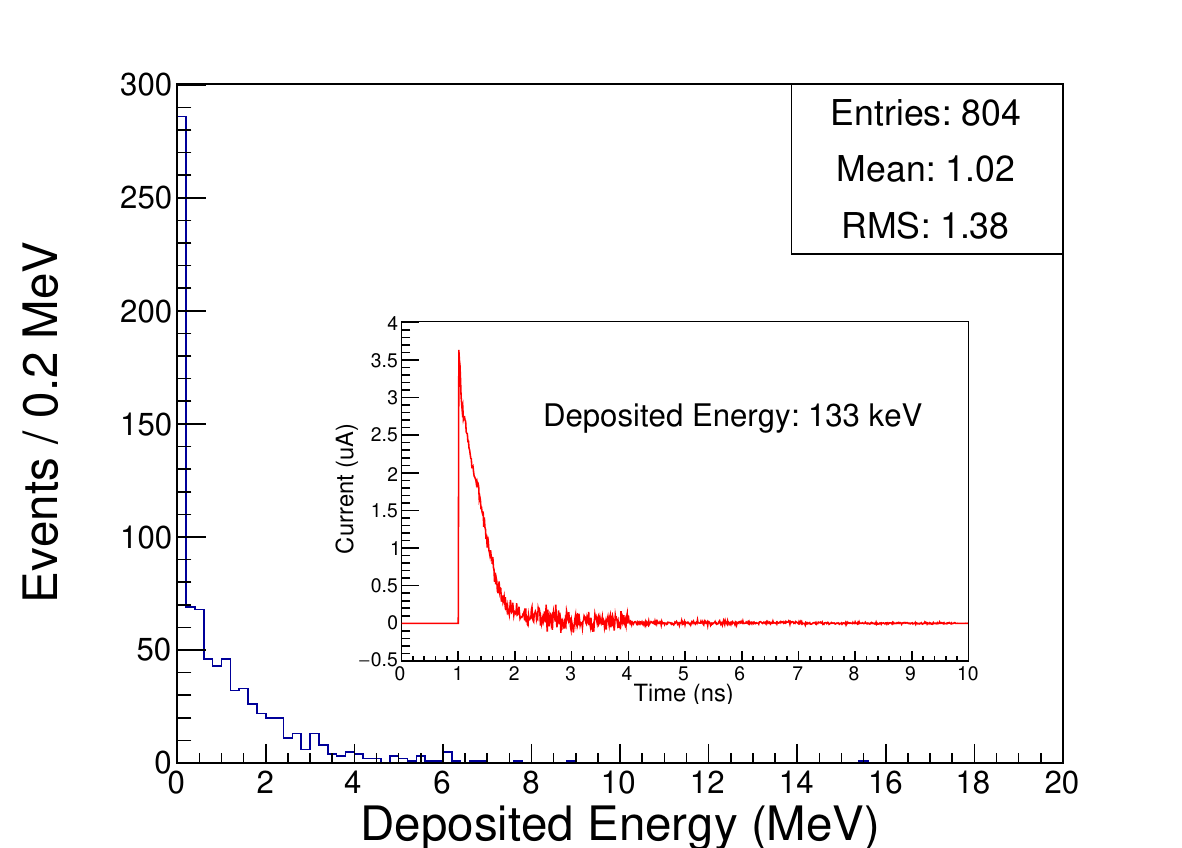}
    \caption{The energy deposition induced by secondary particles generated per bunch crossing within the detector at the -31~mm plane. The inset illustrates the current signal generated by a single pixel.}
    \label{fig:Edep_I}
\end{figure}

As previously stated, when the beam is properly aligned, the optimised detector layout can meet the design requirements of CEPC, Figure~\ref{fig:Edep_I} presents the energy deposition of secondary particles generated per bunch crossing across the entire detector at the -31~mm detector plane under normal luminosity conditions, along with the resultant current signal generated by a single pixel within the entire array, recorded within a 1~ms time frame. As shown in Figure~\ref{fig:Edep_I}, the total energy deposition across the entire detector is about 1~MeV, while a single pixel within the array generates a current pulse characterized by $\mu$A-level amplitude and 3~ns duration. Similarly, detector positioned at the 31~mm detector plane exhibits analogous behavior; therefore it is not displayed.

\begin{figure}[h]
    \centering
    \subfigure[]{ \label{fig:response at L0}
    \includegraphics[scale=0.65]{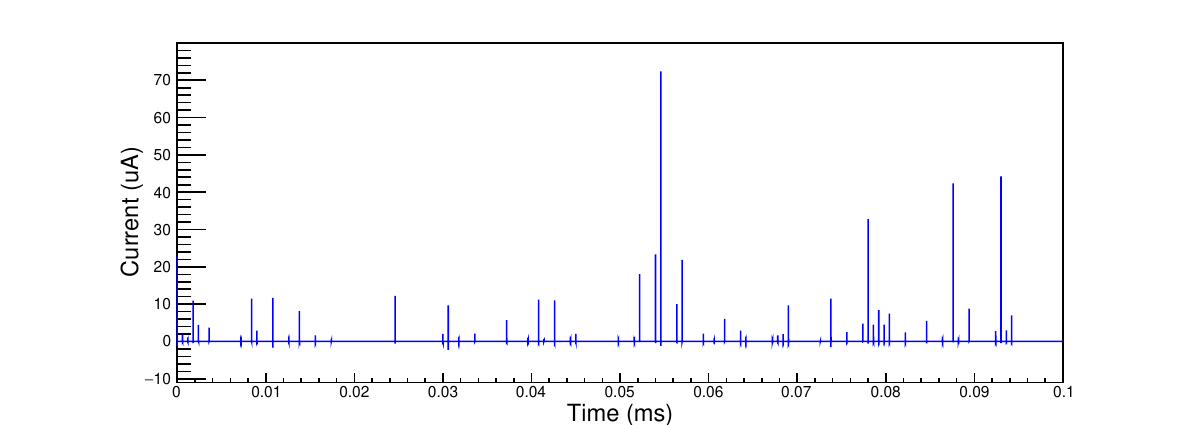}}\\  
    \subfigure[]{ \label{fig:response at 0.1L0}
    \includegraphics[scale=0.65]{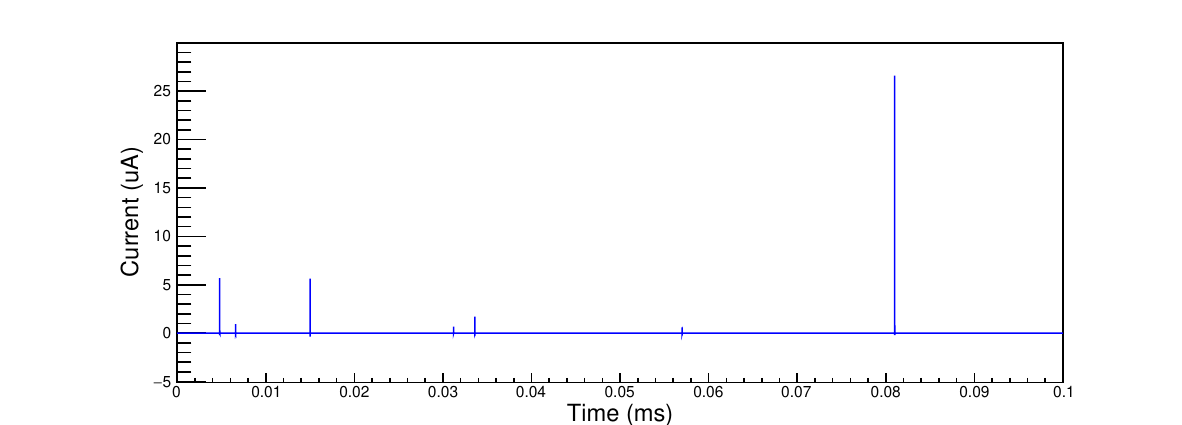}}
    \caption{Current signal response of a channel at (a) nominal luminosity and (b) 10\% nominal luminosity within 0.1 ms}
    \label{fig:response of different L}
\end{figure}

\begin{figure}[h]
    \centering
    \subfigure[]{ \label{fig:Amp_dis}
    \includegraphics[scale=0.3]{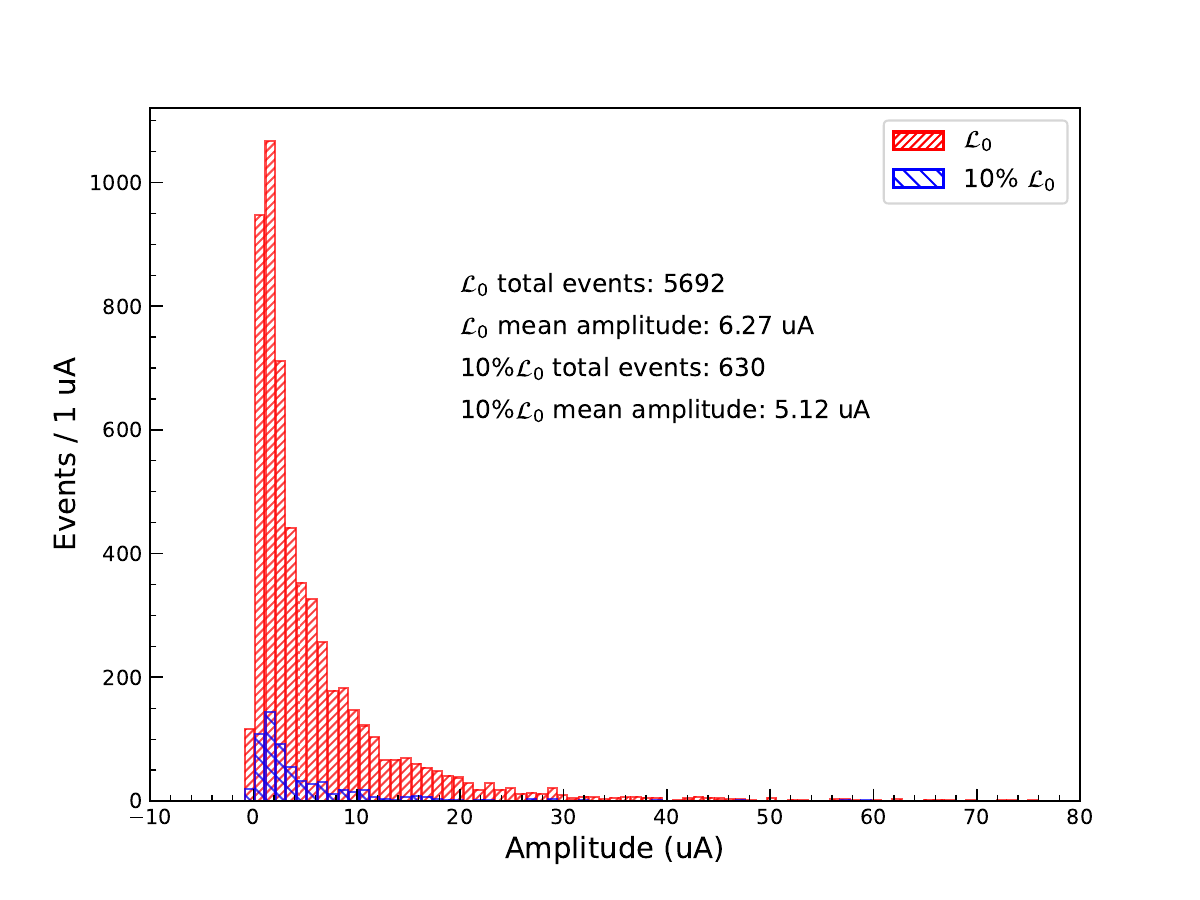}}  
    \subfigure[]{ \label{fig:TSC_Luminosity}
    \includegraphics[scale=0.3]{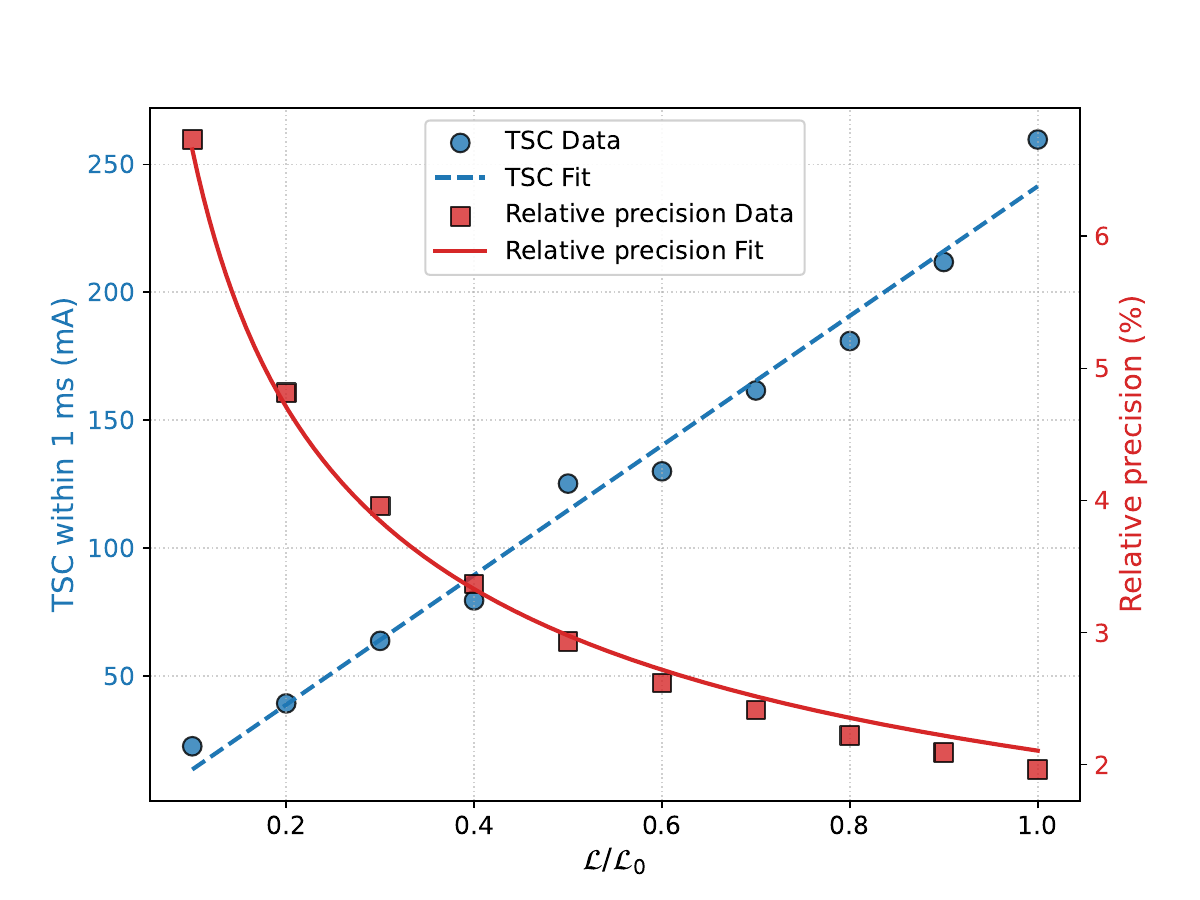}}
    \caption{(a) The amplitude distribution of two entire detectors at nominal luminosity $\mathcal{L}_{0}$ and 10\% $\mathcal{L}_{0}$ and (b) The relationship between the TSC within 1 ms and the ratio of actual luminosity to nominal luminosity}
\end{figure}

However, as a fast luminosity detector, it is imperative for the device to respond swiftly to fluctuations in luminosity. Consequently, simulating its response across various luminosity levels is essential. Figure~\ref{fig:response of different L} illustrates the signal conditions for a channel at luminosity levels of $\mathcal{L}_{0}$(nominal luminosity of Higgs mode, $5 \times 10^{34} \, \text{cm}^{-2}\text{s}^{-1}$) (Figure~\ref{fig:response at L0}) and 10\% of $\mathcal{L}_{0}$ (Figure~\ref{fig:response at 0.1L0}) within 0.1~ms. Figure~\ref{fig:response of different L} illustrates that a reduction in luminosity is associated with a decrease in the number of signal peaks. Figure~\ref{fig:Amp_dis} illustrates the amplitude distribution of the signal over two entire detectors within 1 ms at nominal luminosity $\mathcal{L}_{0}$ and 10\% $\mathcal{L}_{0}$. It was observed that upon reducing the luminosity to one-tenth of its nominal value, the total number of peaks produced by two detectors diminished to 11\% of the initial count, and the average peak value decreased to 76\% of its original magnitude.

Assuming the data acquisition system samples at the frequency of 10~GHz, to accurately describe the correspondence between the signal response and the luminosity, as the RAW SUM signal used in the SuperKEKB fast luminosity detector design\cite{pang:tel-03092297}, TSC is introduced:

\begin{equation}
    \mathrm{TSC} = \sum_{j=1}^{2} \, \sum_{s \in \mathcal{S}} \, \sum_{p \in \mathcal{P}} i_j(s, p)
    \label{TSC}
\end{equation}

In the eq.(\ref{TSC}), $p$ denotes the channel associated with each pixel within the single-sided detector of the beam pipe. The summation is conducted across all 12 pixels. $s$ signifies the sampling points within a 1~ms time window, with the summation encompassing all sampling points. Additionally, $j$ represents the signals generated by the detectors situated on both sides of the beam pipe, with the summation including the sampling signals from both detectors. The relationship between the TSC within 1~ms and the ratio of actual luminosity to nominal luminosity, along with the relative precision at different luminosity ratio is depicted in Figure~\ref{fig:TSC_Luminosity}.
Figure~\ref{fig:TSC_Luminosity} reveals a clear linear correlation between the TSC value within 1~ms and the ratio of actual luminosity to nominal luminosity, indicating that TSC within 1~ms is highly sensitive to luminosity variations. The linear relationship confirms that luminosity variations can be monitored through TSC measurements, thereby enabling verification of beam alignment conditions. At the same time, the relationship between relative precision and luminosity is approximately given by $\nu \propto 1/\sqrt{\mathcal{L}}$. As the luminosity decreases, the relative precision will also decrease accordingly.

\section{Summary} \label{sec:summary}

To meet the real-time beam feedback requirements in the horizontal direction for CEPC, we designed a fast luminosity detector based on 4H-SiC with dimensions of 3~cm in the vertical direction and 1~cm along the beam direction. The precision is 2\% under normal luminosity conditions and a noise threshold of 1 $\mu$A per channel within 1~ms, which meets the precision requirements for fast luminosity detectors. Additionally, we considered the detector's response to decreasing luminosity. Simulation results indicate that the TSC value within 1~ms is highly sensitive to luminosity changes, making it suitable for assessing beam state. However, this study does not address the effects of beam-beam deflections on detector signals, which may influence the signals at Position 1 to some extent but have minimal impact at position 3. Future research should include systematic simulations of beam-beam deflections to evaluate their influence on detector signals. At the same time, actual devices and their matched electronic systems should be developed to meet the CEPC's high luminosity requirements as effectively as possible.

\section*{Acknowledgments}
This work is supported by the National Natural Science Foundation of China (Nos.~12205321, 12375184, 12305207, and 12405219), National Key Research and Development Program of China under Grant No.~2023YFA1605902 from the Ministry of Science and Technology, China Postdoctoral Science Foundation (2022M710085), Natural Science Foundation of Shandong Province Youth Fund (ZR2022QA098) under CERN RD50-2023-11 Collaboration framework. We acknowledge the RASER team for their useful discussions as well as by the CNRS International Research Network FCPPN "France-China Particle Physics Network".

\bibliographystyle{elsarticle-num} 
\bibliography{CFLM}






\end{document}